# Small-scale motions in solar filaments as the precursors of eruptions

**Daikichi S**EKI[1,2*]**, Kenichi O**TSUJI,[2]**, Hiroaki I**SOBE[3]**, Takako T. I**SHII[2]**, Kiyoshi I**CHIMOTO[2] **and Kazunari S**HIBATA[2]

[1]Graduate School of Advanced Integrated Studies in Human Survivability, Kyoto University, Sakyo, Kyoto 606-8306, Japan

[2]Astronomical Observatory, Kyoto University, Yamashina, Kyoto, 607-8471, Japan

[3]Faculty of Fine Arts, Kyoto City University of Arts, Nishikyo, Kyoto, 610-1197, Japan

[*]E-mail: seki@kwasan.kyoto-u.ac.jp

**Abstract**

Filaments, the dense cooler plasma floating in the solar corona supported by magnetic fields, generally exhibit certain activations before they erupt. In our previous study (Seki et al. 2017 ), we observed that the standard deviation of the line-of-sight (LOS) velocities of the small- scale motions in a filament increased prior to its eruption. However, because that study only analyzed one event, it is unclear whether such an increase in the standard deviation of LOS velocities is common in filament eruptions.

In this study, 12 filaments that vanished in Hα line center images were analyzed in a manner similar to the one in our previous work; these included two quiescent filaments, four active region filaments, and six intermediate filaments. We verified that in all the 12 events, the standard deviation of the LOS velocities increased before the filaments vanished. Moreover, we observed that the quiescent filaments had approximately 10 times longer duration of an increase in the standard deviation than the other types of filaments. We concluded that the standard deviation of the LOS velocities of the small-scale motions in a filament can potentially be used as the precursor of a filament eruption.

**1 Introduction**

A dark filament, or a prominence, is the dense and cool plasma supported by magnetic fields in the solar corona, with a plasma density of $10^9$–$10^{10}$ cm$^{-3}$ and a temperature of $10^4$ K. In general, a filament is globally stable; however, at the end of its life, it generally becomes unstable and erupts (Parenti 2014). A filament eruption is generally associated with various solar eruptive phenomena such as flare, coronal mass ejections (CMEs), and very large arcade formation on the quiet Sun. Although they are diverse in size, morphology, and emitting radiation spectrum, they are considered to be the different aspects of a common magnetohydrodynamic process that involves plasma ejection and magnetic reconnection (Shibata & Magara 2011).

Filament eruptions are generally preceded by filament activations (Smith & Ramsey 1964; Parenti 2014) such as slow rise (Sterling & Moore 2004; Sterling et al. 2011), twisting and rotational motions (Gosain et al. 2009), fragmentary brightenings (Ofman et al. 1998), weak heating (Chifor et al. 2006 ), oscillatory plasma motions (Isobe & Tripathi 2006), and active internal motions (Tandberg-Hanssen 1995; Seki et al. 2017 ) in the filaments.

In the more general context of solar eruptions, various types of "triggers" have been proposed, including emerging magnetic flux (Feynman & Martin 1995; Chen & Shibata 2000 ; Kusano et al. 2012), magnetic reconnection at various magnetic configurations (Antiochos et al. 1999 ; Moore et al. 2001), and helicity injection (Magara & Tsuneta 2008; Harra et al. 2009). Among others, an increase in the non-thermal velocity prior to the onset of a flare is discussed in Harra et al. 2001. By spectroscopic observation of the coronal lines, Harra et al. 2001 identified a non-thermal line broadening prior to an increase in the X-ray flux and electron temperature, indicating an increase in the turbulent motion occurring prior to the onset of the flare.

In our previous study (Seki et al. 2017 ), we analyzed an intermediate filament near NOAA 12605, which erupted on November 5, 2016. The data were captured by the Solar Dynamics Doppler Imager (SDDI) (Ichimoto et al. 2017) newly installed on the Solar Magnetic Activity Research Telescope (SMART) (UeNo et al. 2004) at the Hida Observatory. The SDDI captures solar full-disk images in wavelengths from Hα - 9.0 A to Hα + 9.0 A at steps of 0.25 A. It permits us to monitor the Hα line profile and thus determine the line-of-sight (LOS) velocity map (such as the bottom panels in Figure 1) prior to and during the eruption of a filament. From the velocity map, we prepared a histogram of the LOS velocity and calculated the standard deviation of the velocity distribution to quantify the small-scale motions in the filament.

We determined that

1. although the standard deviation was almost constant at approximately 2–3 km s$^{-1}$ from 29 to 21 h prior to the eruption, it had increased to 4–5 km s$^{-1}$ 6 h prior to the eruption, whereas the mean LOS velocity was constant at 0 km s$^{-1}$. We also determined that
2. approximately 1 h prior to eruption, the average of the LOS velocity distribution assumed a neg- ative value; this implies that the filament started to move toward the Earth globally, whereas the standard deviation increased to 10 km s$^{-1}$.

The second observation is likely to correspond to the commonly observed slow-rise phase (Sterling & Moore 2004; Isobe & Tripathi 2006); meanwhile, the first observation indicates an increase in the turbulent motion of the filament prior to the onset of the slow-rise phase and may be regarded as an indication that the filament was approaching an unstable state or a loss of equilibrium, which impelled the eruption. Hereafter, we call such a pre-slow-rise period with increasing standard deviation but without significant change in the mean LOS velocity, as "Phase 1." "Phase 2" is defined as the period from the time when both the increase in the standard deviation and the systematic change in the mean LOS velocity start, to the time when the filament vanishes. Phase 1 can be regarded as a gradual increase in the turbulent motion a significant time prior to the onset of the slow rise; Phase 2 includes the slow-rise phase as well as the eruption phase (with a further increase in the turbulent motion).

The objective of this study is to examine whether these two observations of Seki et al. 2017 generally hold for other filament eruptions. For this purpose, we examined 17 filament disappearance events observed by the SDDI from the beginning of its routine observation in May 2016, to May 2017. After removing low-quality data sets, we analyzed 12 filaments, which included two quiescent filaments, four active region filaments, and six intermediate filaments, similarly as in Seki et al. 2017. We present the event list and the method in section 2, and the results are presented in section 3. In section 4, we summarize and discuss our results.

## 2 Observations

### 2.1 Event List

The SDDI installed on the SMART at Hida Observatory has been conducting routine observations since May 1, 2016. It captures solar full-disk images in 73 wavelengths: from Hα line center - 9.0 A to Hα line center + 9.0A at steps of 0.25A ,i.e.,36 positions in the blue wing, Hα line center, and 36 positions in the red wing. A set of the images in the 73 wavelengths is obtained with a time cadence of 15 s and spatial sampling of 1.23 arcsec per pixel (Ichimoto et al. 2017). When the weather permitted, the SDDI could continuously monitor the Sun during the daytime in Hida. The details of the instrument, examples of the obtained images, and spectral profiles are available in Ichimoto et al. 2017.

**Table 1.** Filament disappearance events observed by SDDI (May 2016–May 2017)

| Event | Date[a] | F.D. Time[b] [UT] | Type[c] | Flare Time[d] & Class [UT] | CME[e] [UT] | Analyzed[f] |
|---|---|---|---|---|---|---|
| 1 | 2016 May 4 | 01:20 | AF (NOAA 12541) | 01:20 (B 6.9) | – | yes (Fig. 11) |
| 2 | 2016 May 13 | 22:09 | QF | – | – | no |
| 3 | 2016 May 13 | 22:35 | AF (NOAA 12544) | 22:38 (B 5.5) | – | no |
| 4 | 2016 Jul 7 | 07:58 | AF (NOAA 12561) | 07:55 (C 5.0) | – | yes (Fig. 12) |
| 5 | 2016 Aug 10 | 03:13 | IF (NOAA 12574, 12575) | – | 04:00 | no |
| 6 | 2016 Sep 4 | 03:37 | IF (NOAA 12586) | – | – | yes (Fig. 5) |
| 7 | 2016 Sep 9 | 05:44 | IF (NOAA 12588) | – | – | yes (Fig. 6) |
| 8 | 2016 Sep 9 | 22:27 | AF (NOAA 12588) | 22:29 (B 4.0) | – | yes (Fig. 13) |
| 9 | 2016 Oct 1 | 03:06 | QF | 02:23 (B 3.5) | 02:30 | no |
| 10 | 2016 Nov 4 | 04:47 | QF | – | 08:00 | yes (Fig. 3) |
| 11 | 2016 Nov 5 | 03:40 | IF (NOAA 12605) | 04:30 (B 1.1) | 04:36 | yes (Fig. 7) |
| 12 | 2017 Feb 19 | 05:40 | IF (NOAA 12636) | 05:47 (B 3.1) | – | yes (Fig. 1, 2, 8) |
| 13 | 2017 Apr 2 | 06:29 | IF (NOAA 12644, 12647) | – | – | no |
| 14 | 2017 Apr 23 | 04:52 | AF (NOAA 12651) | – | – | yes (Fig. 14) |
| 15 | 2017 Apr 23 | 05:40 | IF (NOAA 12652) | 05:50 (B 1.7) | 06:00 | yes (Fig. 9) |
| 16 | 2017 Apr 24 | 02:06 | QF | – | 05:34 | yes (Fig. 4) |
| 17 | 2017 Apr 30 | 00:38 | IF (NOAA 12653) | 01:00 (B 3.0) | 02:36 | yes (Fig. 10) |

(a) The date on which the event started. (b) The time when a filament completely vanished in H$\alpha$ line center as observed by the SDDI. (c) The type of filament; AF, QF, and IF imply active region filament, quiescent filament, and intermediate filament, respectively. (d) The peak time and the class of a flare determined by soft X-ray flux in the GOES 1.0–8.0 Å channel. (e) The time when the associated CME was first observed by SOHO/LASCO C2. (f) If not analyzed, it is because of the presence of terrestrial clouds in the images or the location of a filament.

By surveying the daily observation logs from May 1, 2016 to May 31, 2017 and visually seeking filament disappearance events, we obtained 17 events, which are presented in Table 1. Some of the data were not used because it was challenging to conduct quality analysis on data contaminated by terrestrial clouds or because our method cannot be applied to a filament located above or very close to the solar limb. Therefore, five events were eventually omitted (Events 2, 3, 5, 9, and 13 in Table 1). With regard to "CME [UT]" in Table 1, we identified the CME associated with a filament eruption by considering its first appearance time, central position angle, and linear speed reported in the SOHO/LASCO CME catalogue (Gopalswamy et al. 2009). Finally, we obtained 12 filament disappearance events, which included two quiescent filaments, four active region filaments, and six intermediate filaments.

2.2 Analysis

We used Beckers' cloud model (Beckers 1964) to calculate the LOS velocity. By applying the model to the 73 images captured at the multiple wavelengths around Hα, we determined the source function, Doppler width, Doppler shift, and optical depth of each filament; we assumed that the source function is constant along the wavelengths and along the LOS direction and that the line absorption coefficient is a Gaussian (Morimoto & Kurokawa 2003a; Morimoto & Kurokawa 2003b; Morimoto et al. 2010 ; Cabezas et al. 2017; Sakaue et al. 2018 ; Seki et al. 2017 ). Then, the LOS velocity at each pixel was calculated from the Doppler shift. The advantages of the SDDI are the wide wavelength coverage around Hα and the high spectral and temporal resolution; these enable us to obtain the LOS velocity distribution in unprecedented detail. Figure 1 shows examples of the data. These images were obtained on February 19, 2017.

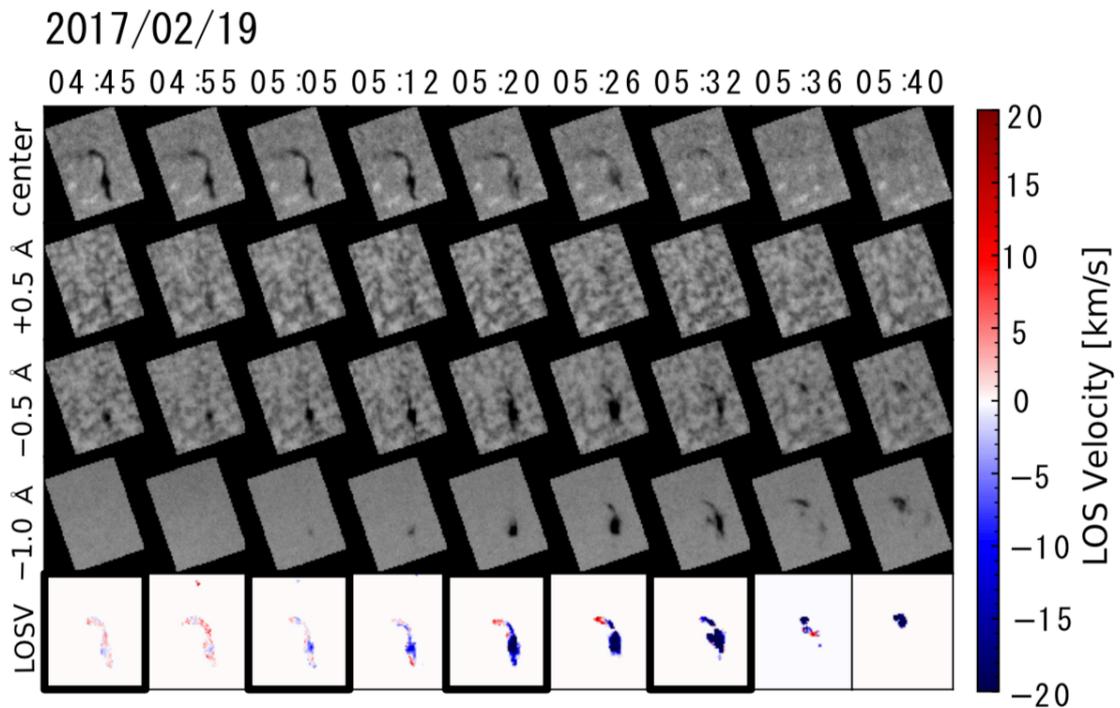

**Fig. 1.** *From top to bottom* : Time series of Hα images at the line center , + 0.5 Å, - 0.5 Å, and - 1.0 Å and of the LOS velocity map of the filament on February 19, 2017 (Event 12). (All the figures in this paper are available in color in the electric version.)

In the following, we explain the procedure of our data analysis in greater detail. It is composed of three steps: development of a mask to obtain the form of a target filament, calculation of the LOS velocity of the filament using the cloud model, and the data selection.

In the first step, we developed a "mask", a binary image that covered an entire target filament. Because Becker's cloud model can be applied only where a "cloud" (= filament) is present above the top of the chromosphere, this process was necessary to determine the pixels where the LOS velocity was to be calculated. We cropped a sub-image from the full-disk image that covered the entire filament prior to and during disappearance. Then, we determined the positions of the pixels where the intensities were lower than $I(\lambda) - 2\sigma_{I(\lambda)}$; here, $\lambda$ is one of the 73 wavelengths, $I(\lambda)$ is the average of the intensities inside the sub-image in $\lambda$, and $\sigma_{I(\lambda)}$ is their standard deviation. This position-determination procedure was conducted for all the 73 wavelengths. Bringing all the positions together, we finally obtained a binary image whose pixels at the same position have one and the other pixels have zero. However, the mask image developed in this manner covers other dark features such as spicules as well as the main body of the filament. To remove such noises, we applied a standard image processing called "erosion and dilation." Dilation is a process wherein if at least one of the surrounding pixels is one for a pixel, it will be set to one. That is, a pixel is set to zero only if all the eight pixels around it are zero. Erosion is the opposite process wherein a pixel is set to one only if all the surrounding eight pixels are one. By executing the dilation process several times after several erosion processes (e.g., executing erosion–erosion–erosion–dilation–dilation in order), we obtained a clean mask image covering only most of the target filament. The number of repetitions, which was determined by trial and error, was different for different events.

After the mask image was produced, we proceeded to the second step—calculation of the LOS velocity of the filament using the cloud model. All the 73 images in the different channels were multiplied by the binary-mask image, and Becker's cloud model was applied to the nonzero pixels. This step yielded the images of the source function, the Doppler width, the Doppler shift, and the optical depth of the filament at a specified time. We applied these two steps for all the data to obtain a time series of these four physical parameters. The Doppler shift was converted to the LOS velocity; in the following, we use only the LOS velocity.

After carrying out these two steps, as the third step, we manually removed the data unsuitable for our analysis. Terrestrial clouds sometimes covered the Sun, resulting in incorrect calculation of the Doppler shift. Because we need to deduce the standard deviation and the average of the LOS velocity, these can significantly affect our result. In addition to this, small high-speed features that were apparently not associated to the filament were occasionally included in the mask and also significantly affected the average and standard deviation calculation. Therefore, we visually examined all the Hα line center images and the LOS velocity maps and removed the data contaminated by terrestrial clouds and small high-speed features, from our analysis.

To quantify the small-scale-plasma motion in each filament, we calculated the standard deviation and average of the LOS velocity distribution. The standard deviation can be regarded as a measure of the enhancement of the turbulent small-scale-plasma motion. Figure 2 shows four representative snapshots of the filament on 2017 February 19, manifesting the increase in standard deviation with time owing to the filament activation and eruption.

**3 Result**

3.1 Quiescent Filaments

3.1.1 Event 10

Figure 3 shows snapshots of Event 10 in Hα line center and SDO/AIA 304 (Lemen et al. 2012); it also shows the time evolution of the average and standard deviation of the LOS velocity of a filament. This filament was a typical large quiescent filament without active regions in its vicinity. It gradually erupted to the solar west and was accompanied by a slow CME with a linear speed of 147 km s$^{-1}$ as revealed by the SOHO/LASCO C2/3. Approximately 45 h prior to the eruption, the standard deviation was approximately 2–3 km s$^{-1}$; it marginally increased to 3–4 km s$^{-1}$ 17 h later. Approximately 5 h prior to eruption, it was about 4 km s$^{-1}$ (Phase 1). From 01:24 UT, the standard deviation started to increase to approximately 5 km s$^{-1}$ and the mean velocity became marginally negative; this manifested the start of the slow-rise phase, i.e., Phase 2. Finally, it disappeared at approximately 04:47 UT (dash-dotted line).

Note that even after the eruption, both the standard deviation and average of the LOS velocity exhibited the values in the right panel of Figure 7. This originated from the statistical summary of the LOS velocity of patchy features such as remnants of the filament and spicules. Although our masking method described in section 2.2 can remove most of the noises originating from them, marginal amounts of such features remained. As stated in section 2.2, our method cannot distinguish the target plasma from other dark features in Hα line center and its wings. However, compared to the main body of the filament, the total amounts of such features were so small prior to and during the eruption that it is highly likely that our result was negligibly affected.

3.1.2 Event 16

Figure 4 shows snapshots of Event 16 in Hα line center and SDO/AIA 304; it also shows the time evolution of the average and standard deviation of the LOS velocity of a filament. This filament was a typical large quiescent filament without active regions in its vicinity; it erupted on April 24, 2017 02:06 UT to the solar north-east, accompanied by a rapid dynamic CME with a linear speed of 854 km s$^{-1}$. There was no notable geomagnetic storm within a few days after the eruption. Approximately 23 h prior to eruption, the standard deviation was 2–3 km s$^{-1}$; from 03:30 UT on April 23, it started to increase to approximately 4 km s$^{-1}$, i.e., Phase 1 started. The increases in both the standard deviation and average were observed from 23:00 UT; however, both the values started to decrease in approximately 30 min. The standard deviation increased again from approximately 01:30 UT on April 24; finally, the filament completely vanished at 02:06 UT.

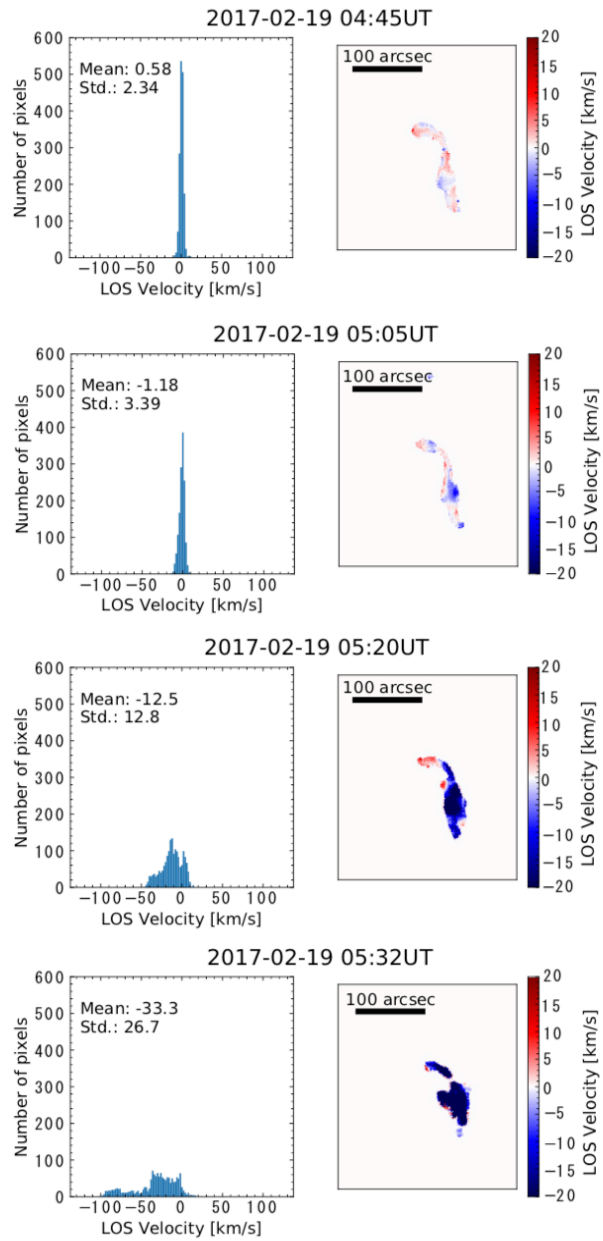

Fig. 2. *Left* : Histograms of the LOS velocity images. Each histogram corresponds to the right image. The mean and standard deviation of the LOS velocity are written on the upper left. Each bin represents 2 km s$^{-1}$. *Right* : Four LOS velocity images inside the black rectangles of Fig.1. Note that the LOS velocity map on each right panel is shown with a scale of lower and upper limits of $\pm$ 20 km s$^{-1}$.

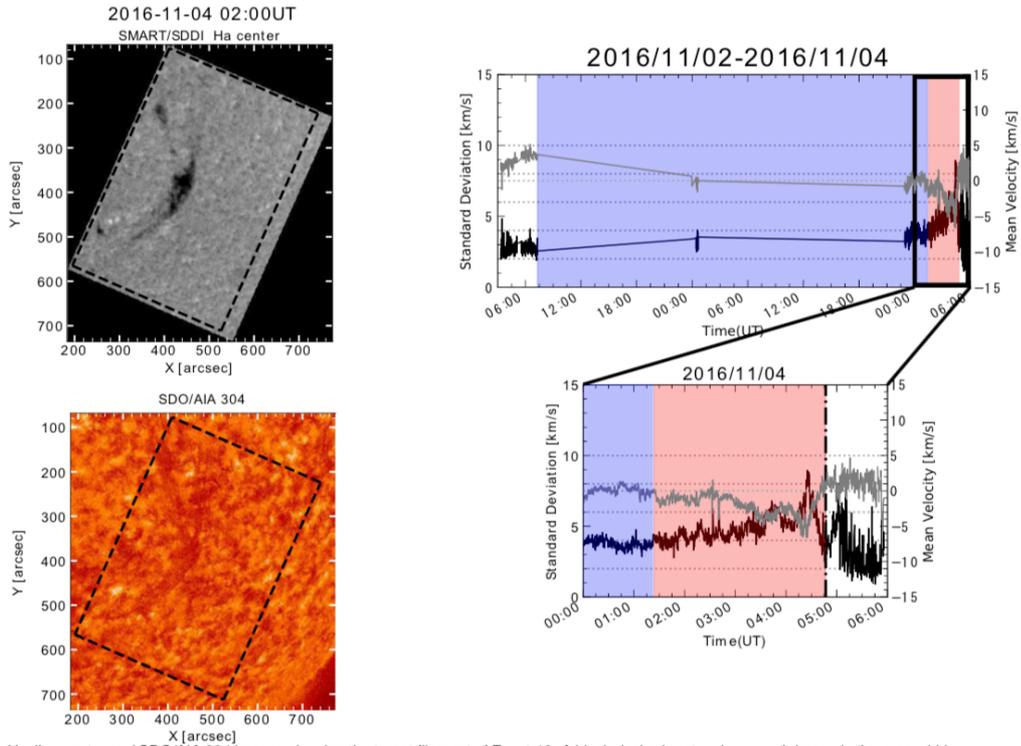

**Fig. 3.** *Left*: Hα line center and SDO/AIA 304 images showing the target filament of Event 10. A black dashed rectangle on each image is the area within which we determined the standard deviation and mean of the LOS velocity. *Right*: Standard deviation (black line, left axis) and average (gray line, right axis) of LOS velocity of filament. The bottom panel is an enlargement of the black rectangle in the top panel. The vertical dash-dotted line in the bottom panel is the time when the filament vanished completely in Hα line center. The blue and red shaded areas correspond to Phases 1 and 2, respectively. The horizontal dotted lines are shown where the standard deviations are 2, 4, 6, 8, and 10 km s$^{-1}$, and the average is 0 km s$^{-1}$.

The increases and decreases in both the values 2.5 h prior to eruption were also observed in our previous work (see Fig. 5 in Seki et al. 2017 ). It was challenging to verify Phase 2 in this case. The possible reason for the absence of Phase 2 is that this filament moved and erupted in an almost perpendicular direction to the LOS direction.

3.2 Intermediate Filaments

3.2.1 Event 6

Figure 5 shows snapshots of Event 6 in Hα line center and SDO/AIA 304; it also shows the time evolution of the average and standard deviation of the LOS velocity of a filament. This filament was located in the vicinity of the active region NOAA 12586 and not on it; therefore, we classified it as an intermediate filament. This filament disappeared on September 4, 2016 03:37 UT in both Hα line center and SDO/AIA 304; moreover, a possible CME associated with this event was not observed by SOHO/LASCO C2/3. At approximately 21:00 UT on September 2, 2016 (approximately 31 h prior to eruption), the standard deviation was approximately 4 km s$^{-1}$. From approximately 00:00 UT on September 4 (approximately 4 h prior to eruption), both the standard deviation and average of the LOS velocity started to change; the former increased to above 4–5 km s$^{-1}$ with a large fluctuation until the eruption, whereas the latter recovered to 0 km s$^{-1}$ approximately 1 h later. Then, this filament vanished at 03:37 UT. There was no signature of Phase 1 in this case. It might be possible that Phase 1 had already occurred before approximately 09:00 UT on September 2, 2016.

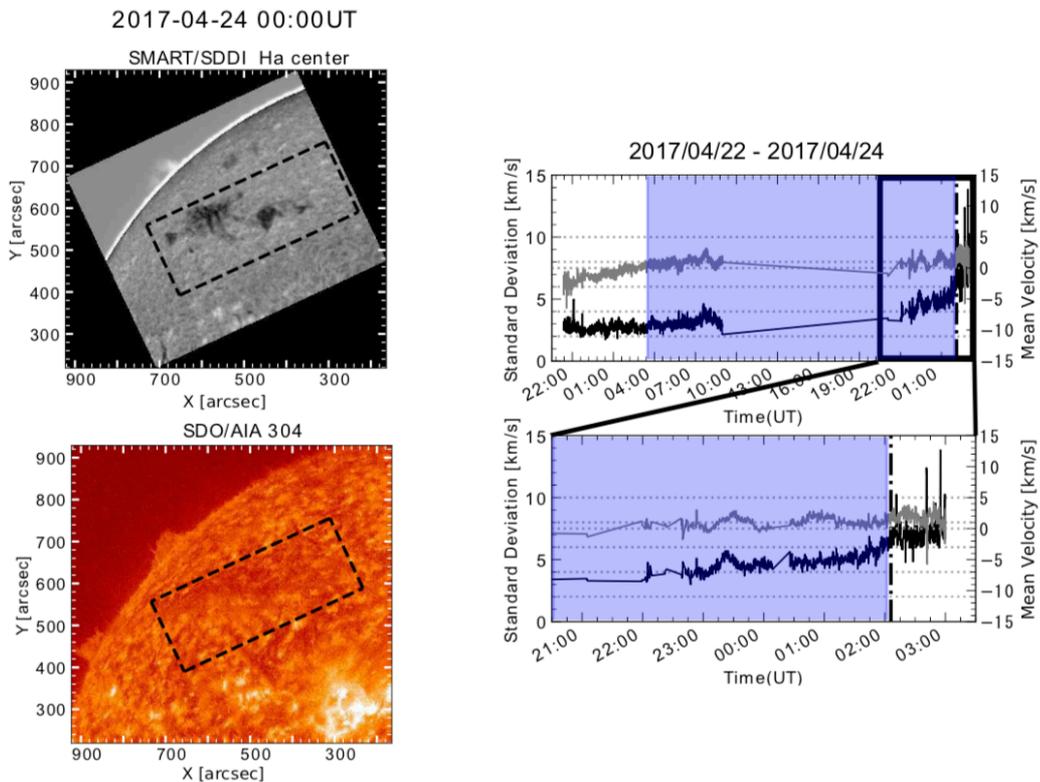

**Fig. 4.** *Left*: Hα line center and SDO/AIA 304 images showing the target filament of Event 16. The definition of the black dashed rectangles is identical to that in Figure 3. *Right*: Standard deviation (black line, left axis) and average (gray line, right axis) of LOS velocity of filament. The bottom panel is an enlargement of the black rectangle in the top panel. The definitions of the blue shaded area, vertical black dash-dotted line, and horizontal gray dotted lines are identical to those in Figure 3. In this case, we could not clearly identify a Phase-2 period (see text), and therefore, red shaded area is absent.

With regard to Phase 2 (red shaded area in Figure 5), except the difference in sign, the profile was similar to that of Events 11 (Seki et al. 2017 ) and 16. Furthermore, in a few wavelengths of the SDO including 304, 193, and 211, this filament displayed an upward motion, i.e., moved toward the solar west in the radial direction; this coincided with the dynamic changes in both the statistical values. This coincidence is consistent with our interpretation of Phase 2, including the slow-rise phase. Note that there was a data gap from 02:55 UT to 03:37 UT on September 4, 2016 because of unfavorable weather condition.

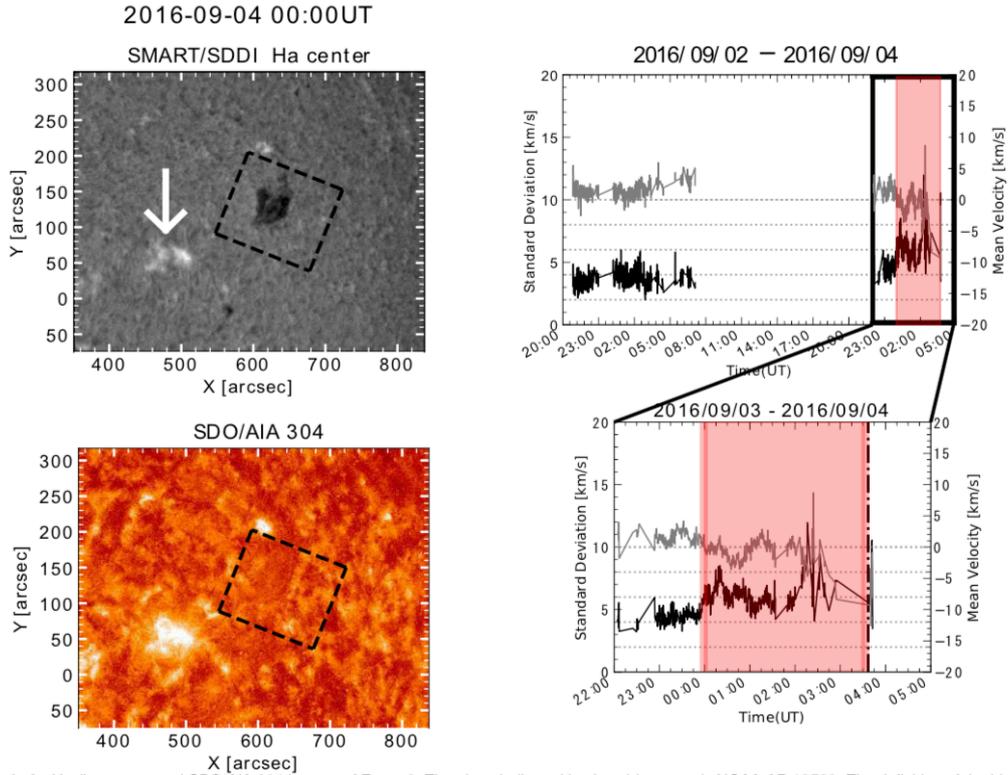

Fig. 5. *Left*: Hα line center and SDO/AIA 304 images of Event 6. The plage indicated by the white arrow is NOAA AR 12588. The definition of the black dashed rectangles is identical to that in Figure 3. *Right*: Standard deviation (black line, left axis) and average (gray line, right axis) of LOS velocity of the filament. The bottom panel is an enlargement of the black rectangle in the top panel. The definitions of the red shaded area, vertical black dash-dotted line, and horizontal gray dotted lines are similar to those in Figure 3. In this case, we could not clearly identify a Phase-1 period (see text); therefore, blue shaded area is absent.

3.2.2 Event 7

Figure 6 shows snapshots of Event 7 in Hα line center and SDO/AIA 304; it also shows the time evolution of the average and standard deviation of the LOS velocity of a filament. This filament was located in the vicinity of the active region NOAA 12588 and vanished on September 9, 2016 in both Hα line center and the SDO/AIA 304. A possible accompanying CME was not observed. From 03:00 UT to 03:45 UT, the standard deviation of the LOS velocity was almost constant at 2 km s$^{-1}$, and the average was also almost constant. Then, only the standard deviation started to increase to 3–4 km s$^{-1}$ (Phase 1) from 03:45 UT to 05:00 UT, whereas the mean LOS velocity was still almost constant at approximately 2 km s$^{-1}$. From 05:10 UT, the filament started to vanish, and at 05:44 UT, it disappeared almost entirely.

Approximately 2.5 h prior to disappearance (approximately 03:14 UT), a B7.9 class flare was observed in NOAA AR 12588; then, the disruption in the loops approximately located in the right- half of the dashed box in Figure 6 was observed at the SDO/AIA 304, 171, 193, and 211. It might be possible that the preceding flare or the disruption was related to the disappearance of the filament.

3.2.3 Event 11

Figure 7 shows snapshots of Event 11 in Hα line center and SDO/AIA 304; it also shows the time evolution of the average and standard deviation of the LOS velocity of a filament. This filament was located in the vicinity of the active region NOAA 12605. It erupted dynamically to the solar north- east on November 5, 2016 04:00 UT, accompanied by a moderate CME with a linear speed of 403 km s$^{-1}$. This CME probably caused the moderate geomagnetic disturbance during November 9–10, 2016. A two-ribbon flare was observed in both Hα line center and SDO/AIA 304, in the vicinity of the filament location immediately after the eruption; this was accompanied by a B1.1-class event in the GOES soft X-ray, which peaked at 04:30 UT. For further details, refer to Seki et al. 2017 . The standard deviation was approximately 2–3 km s$^{-1}$ until approximately 07:00 UT on November 4, 2016; however, at 22:00 UT on November 5, 2016, it was 4–5 km s$^{-1}$, with an average velocity of 0 km s$^{-1}$. Therefore, Phase 1 started during the data-gap period. It gradually increased until 02:30 UT; then, both the standard deviation and mean of LOS velocity started to change more dramatically, i.e., Phase 2 started.

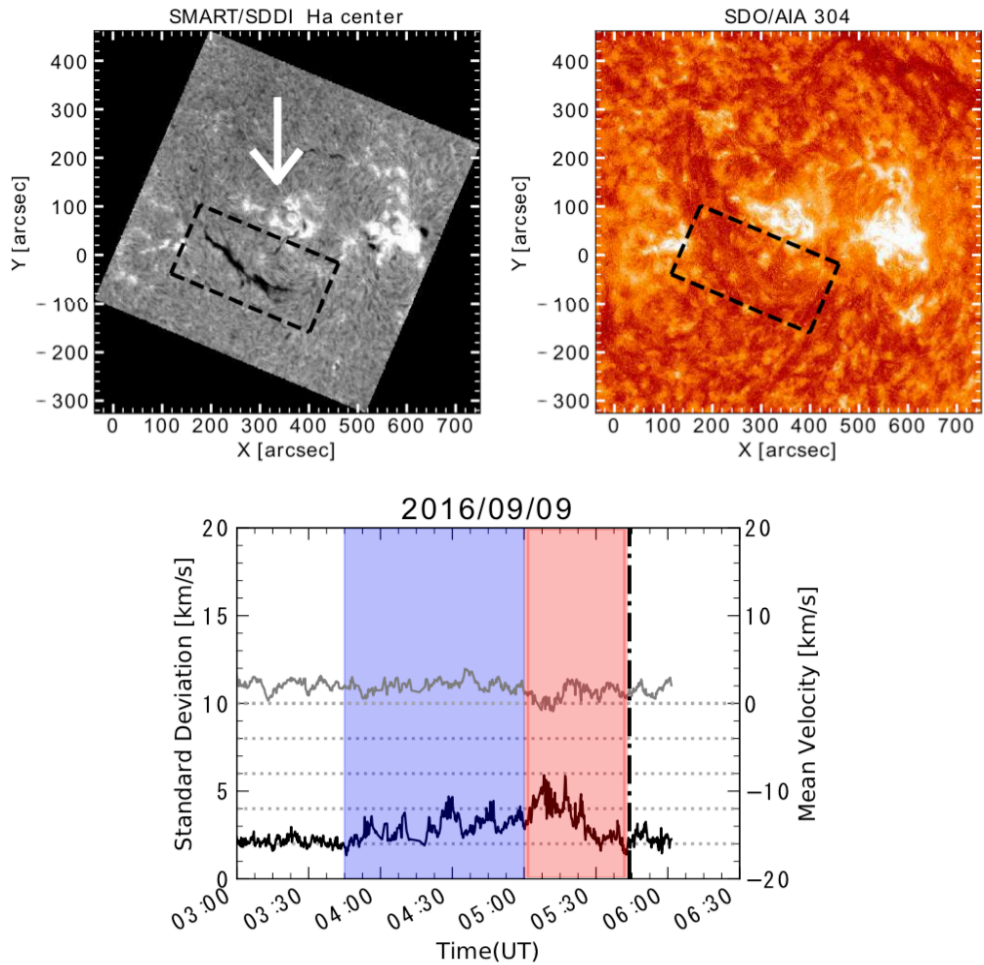

Fig. 6. *Top*: Hα line center and SDO/AIA 304 images of Event 7. The plage indicated by the white arrow is NOAA AR 12588. The definition of black dashed rectangles is identical to that in Figure 3. *Bottom*: Standard deviation (black line, left axis) and average (gray line, right axis) of the LOS velocity of the filament. The definitions of the blue and red shaded areas, vertical black dash-dotted line, and horizontal gray dotted lines are identical to those in Figure 3.

3.2.4 Event 12

Figure 8 shows snapshots of Event 12 in Hα line center and SDO/AIA 304; it also shows the time evolution of the average and standard deviation of the LOS velocity of a filament. This filament was located in the vicinity of the active region NOAA 12636. It erupted rapidly to the solar north on February 19, 2017; however, a possible CME associated with this event was not observed by SOHO/LASCO C2/3. A two-ribbon flare was observed in the vicinity of the filament location immediately after eruption; this was accompanied by a B3.1-class event in the GOES soft X-ray, which peaked at 05:47 UT. From 04:40 UT to 05:10 UT, the standard deviation gradually increased from 2–3 km s$^{-1}$ to 4–5 km s$^{-1}$. The mean LOS velocity was almost constant at approximately 0 km s$^{-1}$ during this period; thus, this increase represents Phase 1. Then, the standard deviation increased more sharply than before, and the average LOS velocity exhibited the gradual decrease. This phase corresponds to Phase 2. From approximately 05:28 UT, both values started to change more dramatically, and finally, it erupted.

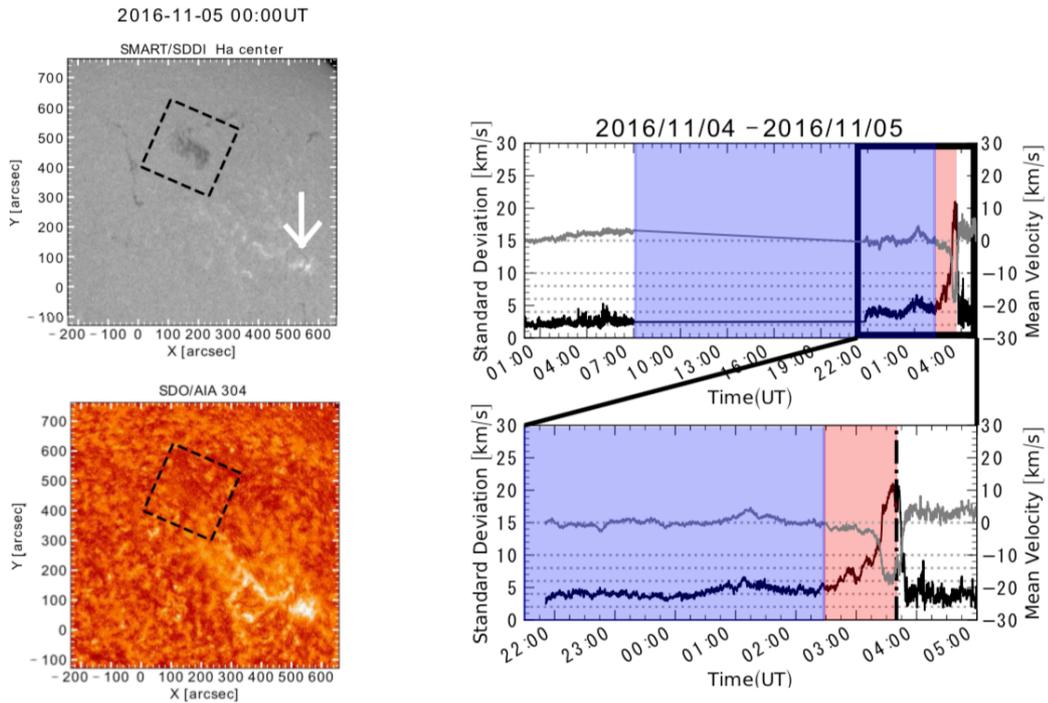

Fig. 7. *Left*: Hα line center and SDO/AIA 304 images of Event 11. The plage indicated by the white arrow is NOAA AR 12605. The definition of black dashed rectangles is identical to that in Figure 3. *Right*: Standard deviation (black line, left axis) and average (gray line, right axis) of LOS velocity of the filament. The bottom panel is an enlargement of the black rectangle in the top panel. The definitions of the blue and red shaded areas, vertical black dash-dotted line, and horizontal gray dotted lines are identical to those in Figure 3.

3.2.5 Event 15

Figure 9 shows snapshots of Event 15 in Hα line center and SDO/AIA 304; it also shows the time evolution of the average and standard deviation of the LOS velocity of a filament. This filament was located in the vicinity of the active region NOAA 12652. It erupted rapidly to the solar south-east on April 23, 2017 accompanied by a rapid and dynamic CME with a linear speed of 955 km s$^{-1}$. A two- ribbon flare was observed in both Hα line center and SDO/AIA 304, in the vicinity of the filament location immediately after eruption; this was accompanied by a B1.7-class event in the GOES soft X-ray, which peaked at 05:50 UT. From 04:15 UT to 04:51 UT (the blue shaded area in Figure 9), the standard deviation gradually increased from 2–3 km s$^{-1}$ to 4–5 km s$^{-1}$. Because the mean LOS velocity was almost constant at approximately 0 km s$^{-1}$ during this period, the period represents Phase 1. Then, the standard deviation increased more sharply than before, and the average of the LOS velocity exhibited the gradual decrease. From approximately 05:18 UT, both values started to change more dramatically, and finally, it erupted.

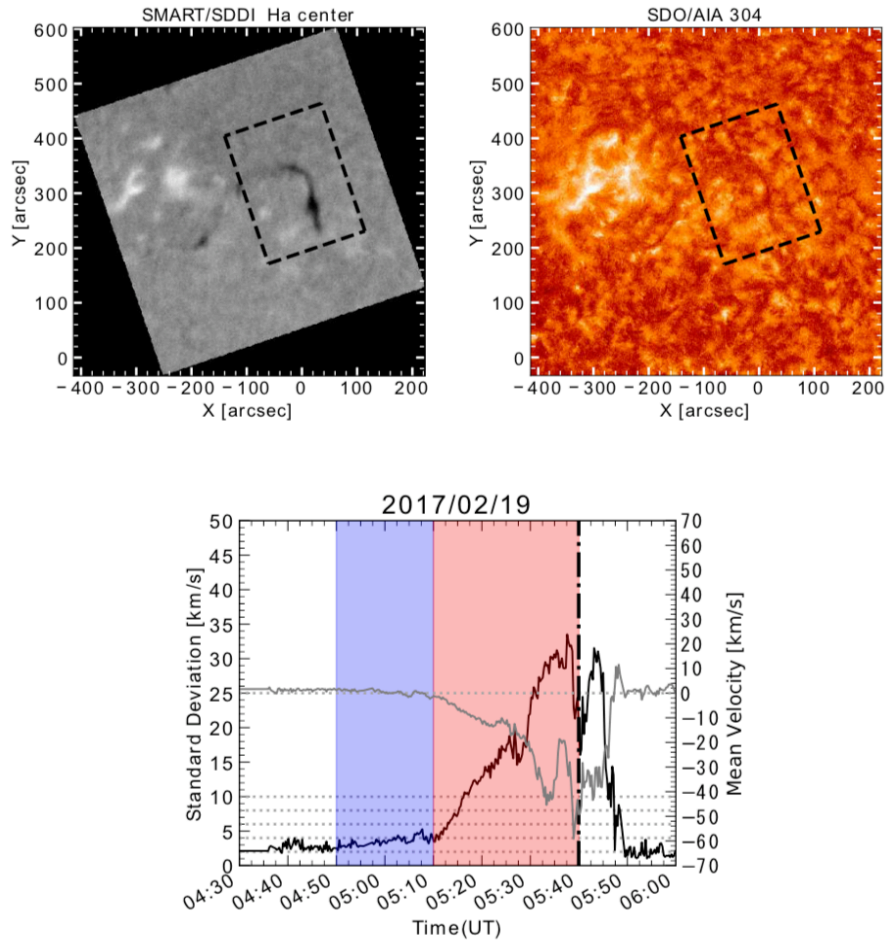

**Fig. 8.** *Top* : Hα line center and SDO/AIA 304 images of Event 12. The plage on the left side of the filament corresponds to NOAA AR 12636. The definition of black dashed rectangles is identical to that in Figure 3. *Bottom* : Standard deviation (black line, left axis) and average (gray line, right axis) of LOS velocity of filament. The definitions of the blue and red shaded areas, vertical black dash-dotted line, and horizontal gray dotted lines are identical to those in Figure 3.

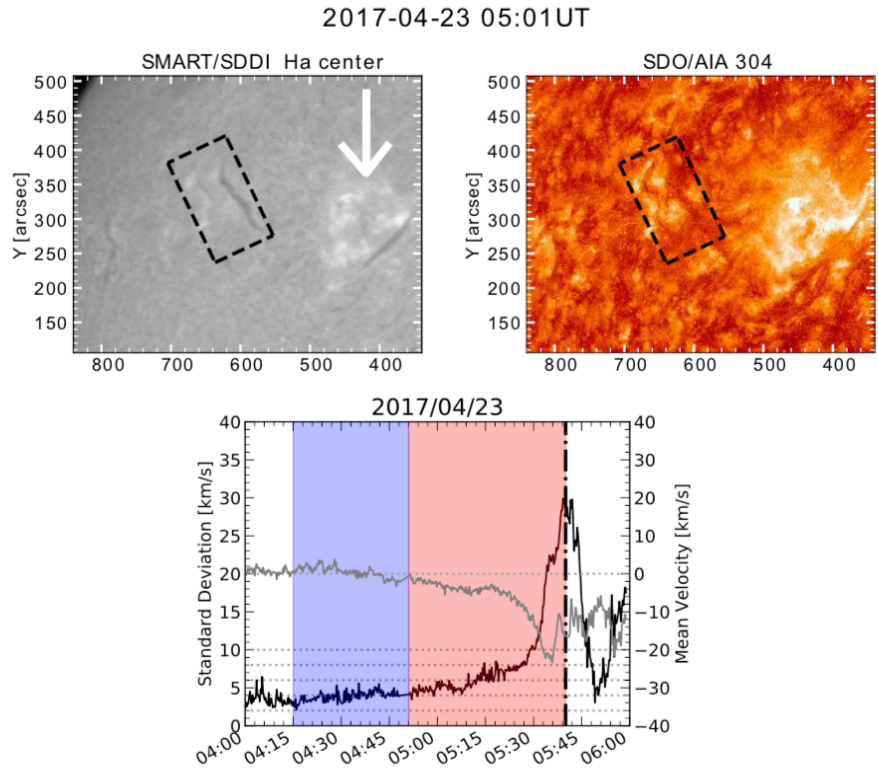

**Fig. 9.** *Top*: Hα line center and SDO/AIA 304 images of Event 15. The plage indicated by the white arrow is NOAA AR 12652. The definition of the black dashed rectangles is identical to that in Figure 3. *Bottom*: Standard deviation (black line, left axis) and average (gray line, right axis) of LOS velocity of filament. The definitions of the blue and red shaded areas, vertical black dash-dotted line, and horizontal gray dotted lines are identical to those in Figure 3.

3.2.6 Event 17

Figure 10 shows snapshots of Event 17 in Hα line center and SDO/AIA 304; it also shows the time evolution of the average and standard deviation of the LOS velocity of a filament. This filament was located in the vicinity of the active region NOAA 12653. It moved dynamically to the solar west and vanished on April 30, 2017, accompanied by a gradual CME with a linear speed of 282 km s$^{-1}$. A two-ribbon flare was observed in both Hα line center and SDO/AIA 304, in the vicinity of the filament location immediately after eruption; this was accompanied by a B3.0-class event in the GOES soft X-ray, which peaked at 01:00 UT. From 23:24 UT to 00:00 UT, the standard deviation gradually increased from 2 km s$^{-1}$ to 4–6 km s$^{-1}$. Because the mean LOS velocity was almost constant during this period, it can be regarded as Phase 1. Then, the standard deviation increased more sharply, and the average of the LOS velocity decreased dramatically.

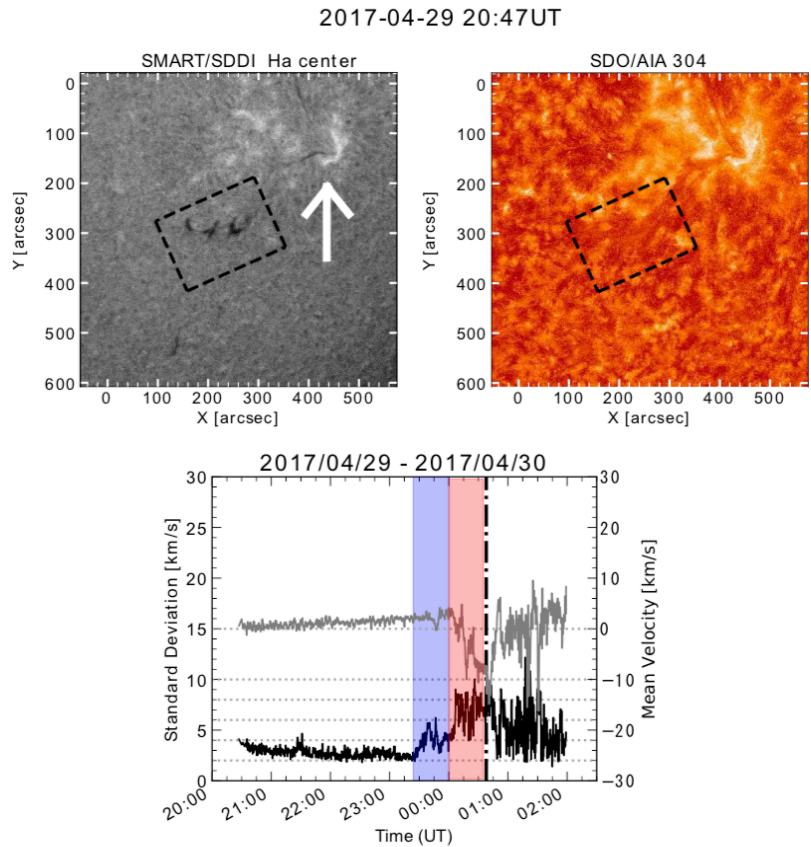

Fig. 10. *Top*: Hα center and SDO/AIA 304 images of Event 17. The plage indicated by the white arrow is NOAA AR 12652. The definition of the black dashed rectangles is identical to that in Figure 3. *Bottom*: Standard deviation (black line, left axis) and average (gray line, right axis) of LOS velocity of filament. The definitions of the blue and red shaded areas, vertical black dash-dotted line, and horizontal gray dotted lines are identical to those in Figure 3.

3.3 Active Region Filaments

We obtained four active-region-filament-disappearance events. In general, the plasma of active region filaments is highly dynamic. This is apparent from the larger amplitudes of both the standard deviation and mean of the LOS velocity compared to those of quiescent and intermediate filaments.

3.3.1 Event 1

Figure 11 shows snapshots of Event 1 in Hα line center and SDO/AIA 304; it also shows the time evolution of the average and standard deviation of the LOS velocity of a filament. This filament was located at NOAA AR 12541. It erupted to the solar west on May 4, 2016; moreover, a possible CME associated with this event was not observed by SOHO/LASCO C2/3. A two-ribbon flare was observed in both Hα line center and SDO/AIA 304, in the vicinity of the filament location immediately after eruption; this was accompanied by a B6.9-class event in the GOES soft X-ray, which peaked at 01:20 UT. From 00:11 UT to 00:20 UT, the standard deviation started to increase from 4 km s$^{-1}$ to 8 km s$^{-1}$, with the mean LOS velocity almost constant (Phase 1). Then, both the statistical values started to proliferate, and the filament erupted at 01:20 UT.

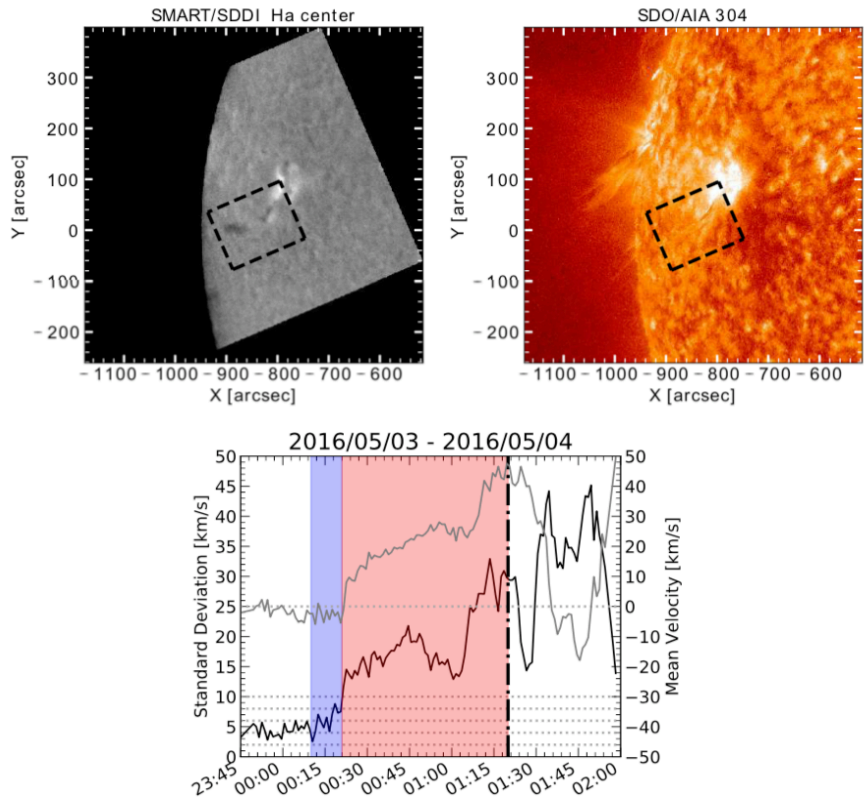

**Fig. 11.** *Top* : Hα line center and SDO/AIA 304 images of Event 1. The active region where the filament is located is NOAA AR 12542. The definition of the black dashed rectangles is identical to that in Figure 3. *Bottom* : Standard deviation (black line, left axis) and average (gray line, right axis) of LOS velocity of filament. The definitions of the blue and red shaded areas, vertical black dash-dotted line, and horizontal gray dotted lines are identical to those in Figure 3.

3.3.2 Event 4

Figure 12 shows snapshots of Event 4 in Hα line center and SDO/AIA 304; it also shows the time evolution of the average and standard deviation of the LOS velocity of a filament. This filament was located at the active region NOAA 12561. This filament erupted to the solar south-west on July 7, 2016; however, a possible CME associated with this event was not observed by SOHO/LASCO C2/3. A two-ribbon flare was observed in the vicinity of the filament location immediately after eruption; this was accompanied by a C5.0-class event in the GOES soft X-ray, which peaked at 07:58 UT. From 07:47 UT, the filament showed the initiation of Phase 2 as the standard deviation started to increase from approximately 6 km s$^{-1}$ to 8–10 km s$^{-1}$, and the average became negative. From 07:52 UT, both the values manifested more dramatic changes, and finally, it erupted.

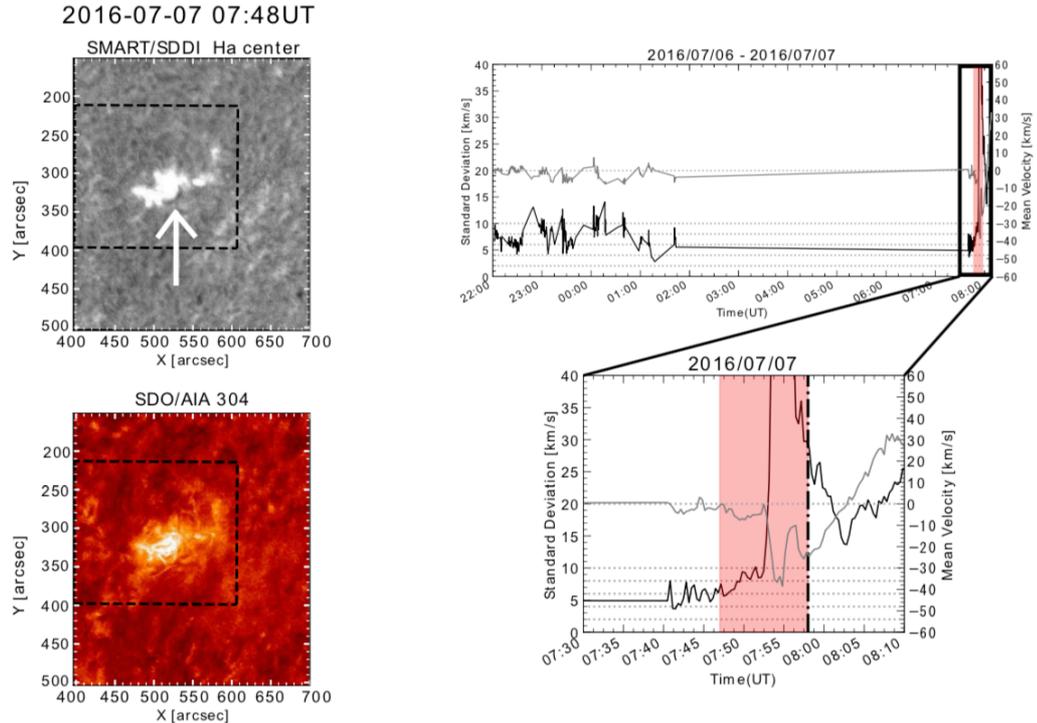

**Fig. 12.** *Left*: Hα line center and SDO/AIA 304 images showing the filament on NOAA AR 12561 (Event 4). The white arrow indicates the target active region filament, and a black rectangle on each image is the area inside which we calculated the standard deviation and mean of LOS velocity. *Right*: Standard deviation (black line, left axis) and average (gray line, right axis) of LOS velocity of filament. The bottom panel is an enlargement of the black rectangle in the top panel. The definitions of the red shaded area, horizontal dotted lines, and vertical dashed and dash-dotted lines are identical to those in Figure 3. In this case, we could not clearly identify a Phase-1 period (see text); therefore, blue shaded area is absent.

It should be noted that this filament vanished in Hα between approximately 02:00 UT and 07:40 UT (the data gap in the upper right panel), and it appeared again at 07:40. Furthermore, Phase 1 was not detected probably because the plasma of the active region filament moves dynamically, and it was unfeasible to recognize the marginal and gradual change in the LOS velocity.

### 3.3.3 Event 8

Figure 13 shows snapshots of Event 8 in Hα line center and SDO/AIA 304; it also shows the time evolution of the average and standard deviation of the LOS velocity of a filament. This filament was located at the active region NOAA 12588. It erupted to the solar north-east on September 9, 2016; however, a possible CME associated with this event was not observed by SOHO/LASCO C2/3. A two-ribbon flare was observed in the vicinity of the filament location immediately after eruption; this was accompanied by a B4.0-class event in the GOES soft X-ray, which peaked at 22:29 UT. From 17 to 14 h prior to eruption, the standard deviation became 4–8 km s$^{-1}$ with large fluctuations. However, 35 min prior to eruption, it became very high ( 17 km s$^{-1}$), whereas the average was approximately 0 km s$^{-1}$ until 22:06 UT. Thus, Phase 1 probably started during the data-gap period. From 22:06 UT, the average of the LOS velocity started to decrease and became negative, whereas the standard deviation remained more or less high; thus, Phase 2 was initiated.

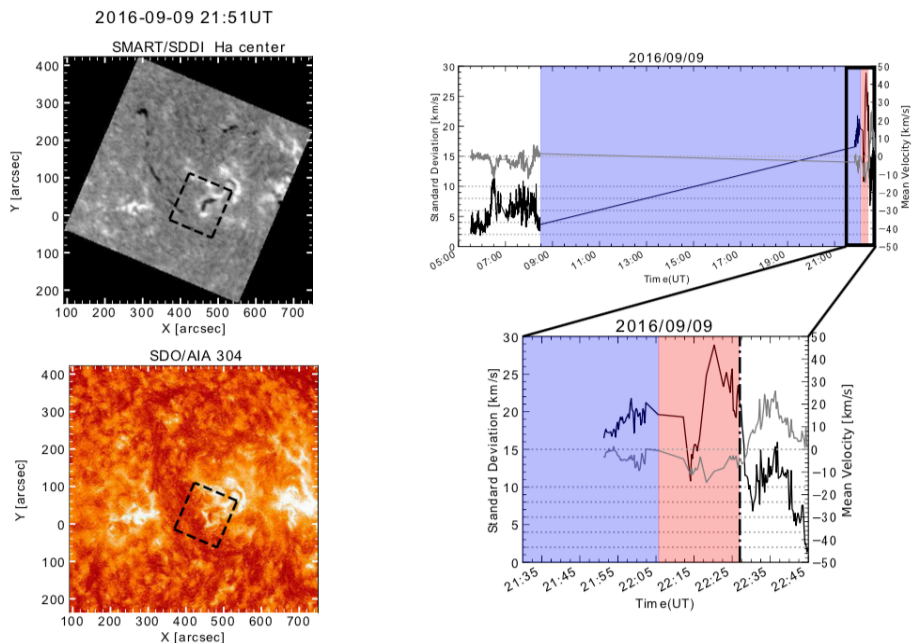

**Fig. 13.** *Left* : Hα line center and SDO/AIA 304 images of the filament on NOAA AR 12588 (Event 8). The definition of the black dashed rectangles is identical to that in Figure 3. *Right* : Standard deviation (black line, left axis) and average (gray line, right axis) of LOS velocity of filament. The bottom panel is an enlargement of the black rectangle in the top panel. The definitions of the blue and red shaded areas, vertical black dash-dotted line, and horizontal gray dotted lines are identical to those in Figure 3.

3.3.4 Event 14

Figure 14 shows snapshots of Event 14 in Hα line center and SDO/AIA 304; it also shows the time evolution of the average and standard deviation of the LOS velocity of a filament. This filament was located at the active region NOAA 12651 and vanished on April 23, 2017; moreover, a possible CME associated with this event was not observed by SOHO/LASCO C2/3. In this event, a B3.8-class flare in the GOES soft X-ray, which peaked at 03:06 UT, was observed at NOAA AR 12651 approximately 1.8 h prior to disappearance. Within a short period after this flare, the filament started to vanish, and the dynamic changes in both the standard deviation and average of the LOS velocity began. In this case, Phase 1 was difficult to confirm.

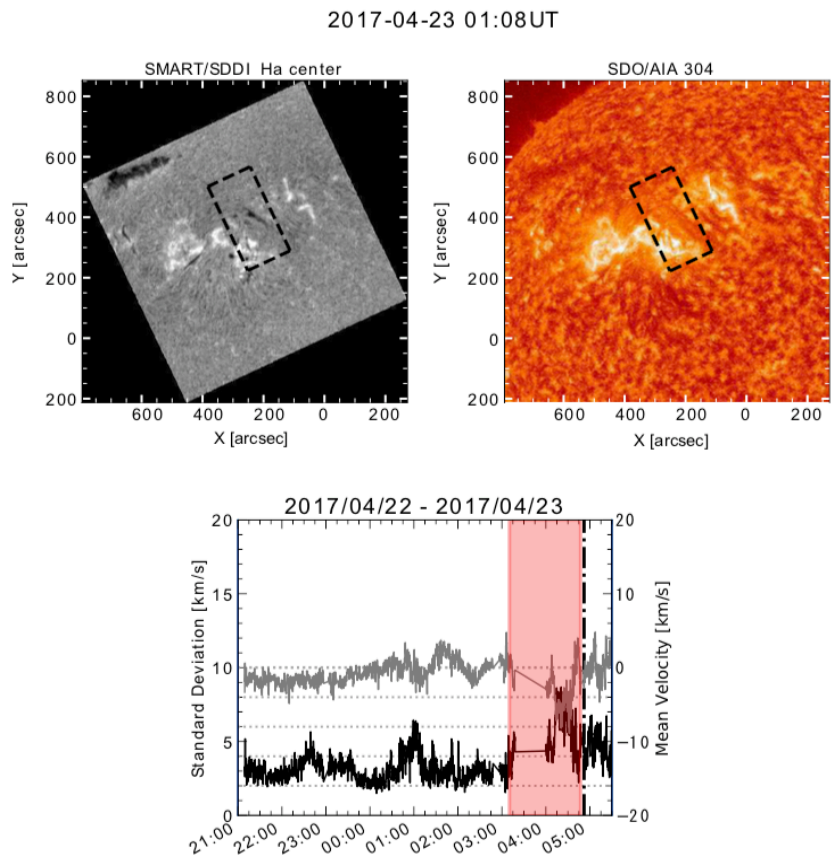

**Fig. 14.** *Top*: Hα line center and SDO/AIA 304 images of Event 14. The active region where the filament was located is NOAA AR 12651. The definition of the black dashed rectangles is identical to that in Figure 3. *Bottom*: Standard deviation (black line, left axis) and average (gray line, right axis) of LOS velocity of filament. The definitions of the red shade area, vertical black dash-dotted line, and horizontal gray dotted lines are identical to those in Figure 3. In this case, we could not clearly identify a Phase-1 period; therefore, blue shaded area is absent.

**4 Summary & Discussion**

In this study, we analyzed 12 filament-disappearance events observed by the SMART/SDDI in Hida Observatory from May 2016 to May 2017 in the same manner as that in Seki et al. 2017 with the purpose of clarifying whether the precursor of a filament eruption indicated in Seki et al. 2017 could be detected as a common feature prior to other filament-disappearance events. Owing to the narrow passband width around Hα line center and the high time cadence of the SDDI, we succeeded in obtaining the unprecedented detailed LOS velocity maps of filaments. Moreover, by tracking the average and standard deviation of the LOS velocity of each filament, we investigated the existence of Phase 1 (the period during which the standard deviation increases while the average is almost constant) and Phase 2 (from the time of initiation of the systematic changes of both the statistical values to the time of a filament eruption). The summary of the results is presented in Table 2.

Table 2. Properties and durations of Phases 1 and 2.

| Event | Type | Phase 1 [hours] | Phase 2 [hours] | CME [UT] | Figure |
|---|---|---|---|---|---|
| 1 | AF (NOAA 12541) | 0.18 | 1.0 | – | Fig. 11 |
| 4 | AF (NOAA 12561) | – | 0.17 | – | Fig. 12 |
| 6 | IF (NOAA 12586) | – | 3.5 | – | Fig. 5 |
| 7 | IF (NOAA 12588) | 1.2 | 0.75 | – | Fig. 6 |
| 8 | AF (NOAA 12588) | 14* | 0.27 | – | Fig. 13 |
| 10 | QF | 25–42 | 3.5 | 08:00 | Fig. 3 |
| 11 | IF (NOAA 12605) | 20* | 1.2 | 04:36 | Fig. 7 |
| 12 | IF (NOAA 12636) | 0.32 | 0.32 | – | Fig. 8 |
| 14 | AF (NOAA 12651) | – | 1.7 | – | Fig. 14 |
| 15 | IF (NOAA 12652) | 0.60 | 0.82 | 06:00 | Fig. 9 |
| 16 | QF | 23 | – | 05:34 | Fig. 4 |
| 17 | IF (NOAA 12653) | 0.60 | 0.63 | 02:36 | Fig. 10 |

* The value with an asterisk (Phase-1 duration of Event 8 or 11) could be overestimated because of the absence of data.

In all the 12 events except Event 16, Phase 2 was detected prior to the disappearance of the filament, regardless of the filament types and whether the disappearance was associated with a CME or not. In addition, it should be noted that in Event 16, the filament moved and erupted perpendic- ularly to the LOS direction in appearance so that the LOS component of its velocity was marginal. Therefore, we conclude from this study that in general, the Phase-2 period is observed prior to fila- ment disappearance.

Now, we present three other findings of this study. First, our results also reveal wide variations in Phase 1 and Phase 2, ranging from 0.18 to 42 h and 0.17 to 3.5 h, respectively. Omitting the possibly overestimated durations of Phase 1 (Events 8 and 11), we observe that a quiescent filament has a longer Phase-1 duration than the other two types of filaments by one or two orders of magnitude. This difference is probably owing to the difference in the Alfven time of a filament, although further study is required to verify this. Meanwhile, although the Phase-2 duration also varies among the events by an order, such a relation between the types of filaments and the duration is not evident.

Secondly, in all the cases of the intermediate and quiescent filaments, the standard deviation during Phase 1 generally changed from 2–3 km s$^{-1}$ to 4–5 km s$^{-1}$. This common transition was observed regardless of the CME association and the size and types of filament. In addition, Kubota & Uesugi 1986 reported that approximately 24 h prior to disappearance, the mean and standard deviation of the LOS velocity of the quiescent filament that disappeared at 04:10 UT on May 8, 1984, were 0.92 km s$^{-1}$ and 2.1 km s$^{-1}$, respectively.

Thirdly, in Events 11 and 16, both the standard deviation and average of the LOS velocity manifested temporary, significant, and systematic increases in a specific period (from 00:30 UT to 01:30 UT in Event 11 and from 23:00 UT to 00:00 UT in Event 16). The similar temporary change in both the statistical values is also evident in Event 6 from 00:00 UT to 01:00 UT; the only exception is that the average of the LOS velocity decreased. The possible interpretation of these temporary disturbances is that they reflect the intermittent disruptions such as emerging flux and magnetic re- connection that contribute to the global evolution of the equilibrium of the magnetic flux system.

In terms of space weather, filament eruptions exhibit a potential risk of disturbing it. McAllister et al. 1996 reported the event of a polar crown filament, which erupted on 1994 April 14. This eruption displayed a very large coronal arcade in a soft X-ray image; finally, a very large geomagnetic storm (Dst ∼ -200 nT) occurred three days after the eruption. Severe geomagnetic storm produced by a quiescent filament eruption is also reported by Cliver et al. 2009 . Joselyn & McIntosh 1981 revealed that 42 geomagnetic storms out of 65 were associated with filament disappearances and argued that a filament disappearance could be used as a useful predictor of geomagnetic storms.

From this perspective, our present study, which demonstrated the general appearance of Phase 2 prior to filament eruptions, proposes a new method to predict filament eruptions by utilizing Phase 2 as an effective precursor. Considering our results, this method enables us to predict filament eruptions 1.3 h on an average prior to onsets. Furthermore, it should be emphasized that this method is currently based only on data captured by ground-based telescopes. Although space-borne data are indispensable for space weather prediction, artificial satellites are vulnerable to space weather effects (National Research Council 2008). In this context, the prediction of solar eruptive phenomena based only on the data of ground-based telescopes as in our method is highly likely to be valuable for supporting the current space weather prediction system.

However, in order to realize the operational prediction, further studies, including on how to distinguish Phase 2 from temporary significant increase in both the statistical values (see the previ- ous paragraph), how to address terrestrial clouds contaminating observed images, and how to speed computation, should be conducted. With regard to the infrastructure, a network of ground-based tele- scopes that monitor Hα line center as well as its blue and red wings, such as CHAIN project (UeNo et al. 2007), can play a significant role in the operation (Seki et al. 2018).

## Acknowledgments

This work was partly supported by JSPS KAKENHI grant numbers JP15H05814, JP16H03955, and JP18J23112. The SOHO/LASCO CME Catalog is generated and maintained at the CDAW Data Center by NASA and The Catholic University of America in cooperation with the Naval Research Laboratory. SOHO is a project of international cooperation between ESA and NASA. This research makes use of SunPy, an open-source and free community-developed solar data analysis package written in Python (The SunPy Community et al. 2015). We express our sincere gratitude to the staff of Hida Observatory for conducting the instrument development and daily observations and to a referee for a number of effective recommendations and comments, which improved the quality of this paper significantly.